\documentclass{ws-procs961x669}            
\usepackage{ws-toc,ws-multind}

\begin{document}

\markboth{J. C. Tan et al.}{The Origin of Supermassive Black Holes from Pop III.1 Seeds}

\wstoc{The Origin of Supermassive Black Holes from Pop III.1 Seeds}{J. C. Tan}

\title{The Origin of Supermassive Black Holes from Pop III.1 Seeds}

\author{Jonathan C. Tan}
\aindx{Tan, J. C.}

\address{Dept. of Space, Earth \& Environment, Chalmers Univ. of Technology, Gothenburg, Sweden,\\
and Dept. of Astronomy, University of Virginia, Charlottesville, VA, USA\\
\email{jctan.astro@gmail.com}}

\author{Jasbir Singh}
\aindx{Singh, J.}
\address{INAF - Astronomical Observatory of Brera, via Brera 28, I-20121 Milan, Italy}

\author{Vieri Cammelli}
\aindx{Cammelli, V.}
\address{University of Trieste \& INAF - AOT, via G.B. Tiepolo 11, I-34143 Trieste, Italy\\
and Dept. of Space, Earth \& Environment, Chalmers Univ. of Technology}

\author{Mahsa Sanati}
\aindx{Sanati, M.}
\address{University of Oxford, Keble Road, Oxford OX1 3RH, UK\\
and Dept. of Space, Earth \& Environment, Chalmers Univ. of Technology}

\author{Maya Petkova}
\aindx{Petkova, M.}
\address{Dept. of Space, Earth \& Environment, Chalmers Univ. of Technology, Gothenburg, Sweden}

\author{Devesh Nandal}
\aindx{Nandal, D.}
\address{Dept. of Astronomy, University of Virginia, Charlottesville, VA, USA}

\author{Pierluigi Monaco}
\aindx{Monaco, P.}
\address{University of Trieste \& INAF - AOT, via G.B. Tiepolo 11, I-34143 Trieste, Italy}

\begin{abstract}
The origin of supermassive black holes (SMBHs) is a key open question
for contemporary astrophysics and cosmology. Here we review the
features of a cosmological model of SMBH formation from Pop III.1
seeds, i.e., remnants of metal-free stars forming in locally-isolated
minihalos, where energy injection from dark matter particle
annihilation alters the structure of the protostar allowing growth to
supermassive scales (Banik et al. 2019; Singh et al. 2023; Cammelli et
al. 2024). The Pop III.1 model explains the paucity of
intermediate-mass black holes (IMBHs) via a characteristic SMBH seed
mass of $\sim10^5\:M_\odot$ that is set by the baryonic content of
minihalos. Ionization feedback from supermassive Pop III.1 stars sets
the cosmic number density of SMBHs to be $n_{\rm SMBH}\lesssim
0.2\:{\rm Mpc}^{-3}$. The model then predicts that all SMBHs form by
$z\sim20$ with a spatial distribution that is initially
unclustered. SMBHs at high redshifts $z\gtrsim7$ should all be single
objects, with SMBH binaries and higher order multiples emerging only
at lower redshifts. We also discuss the implications of this model for
SMBH host galaxy properties, occupation fractions, gravitational wave
emission, cosmic reionization, and the nature of dark matter. These
predictions are compared to latest observational results, especially
from HST, JWST and pulsar timing array observations.
\end{abstract}

\bodymatter

\section{Introduction}

The origin of supermassive black holes (SMBHs) that manifest
themselves as active galactic nuclei (AGN) in the centers of most
large galaxies is one of the most pressing open questions of current
astrophysics and cosmology. How do these cosmic behemoths come into
being? The discovery of SMBHs as $z \sim 7-10$ AGN (e.g., Wang et
al. 2021, Bogd\'an et al. 2024; Maiolino et al. 2024a; Fig. 1) indicates
that at least some SMBHs formed very early, during the so-called
``dark ages'' of the Universe, and did so with initial ``seed'' masses
already in the supermassive regime, i.e., $\sim10^5M_\odot$. While
sustained growth at super-Eddington rates could allow lower-mass
seeds, there is little evidence for such accretion rates in observed
samples of lower-redshift quasars (e.g., Wu \& Shen 2022; Zeltyn et
al. 2024), while the handful of examples of high-$z$ luminous quasars
with Eddington ratios of up to $\sim2$ (Yang et al. 2021) are
consistent with obeying the Eddington limit given measurment
uncertainties. Furthermore, the apparent dearth of intermediate-mass
black holes (IMBHs) (e.g., Reines \& Comastri 2016; Baumgart 2017;
Greene et al. 2020; Mummery \& van Velzen 2024) suggests that a mass
scale of $\sim10^5\:M_\odot$ is a characteristic feature of SMBH
formation and that this is widely separated from the $\sim
10-100\:M_\odot$ black holes that are the remnants of standard stellar
populations.

A wide variety of SMBH formation theories have been proposed (see
reviews by, e.g., Rees 1978; Volonteri et al. 2021). Most, including
formation in dense stellar clusters (e.g., G\"urkan et al. 2004;
Freitag et al. 2006; Schleicher et al. 2022) or the popular ``Direct
Collapse'' from massive ($\sim10^8\:M_\odot$), atomically-cooled,
metal-free, irradiated and/or turbulent halos (e.g., Haehnelt \& Rees
1993; Bromm \& Loeb 2003; Chon et al. 2016; Wise et al. 2019; Latif et
al. 2022), struggle to explain a characteristic mass scale of SMBHs of
$\sim10^5\:M_\odot$, i.e., formation of IMBHs would be expected to be
much more common in these scenarios. Furthermore, these models have
great difficulty in seeding {\it all} SMBHs, i.e., to yield the
numbers seen in the local Universe, i.e., $n_{\rm SMBH} \gtrsim
5 \times 10^{-3}\:{\rm Mpc}^{-3}$ (Vika et al. 2009; BTM19). In the
context of direct collapse black holes (DCBHs), Chon et al. (2016)
performed hydrodynamic (HD) simulations of a $(\sim30\:{\rm Mpc})^3$
volume and estimated a co-moving number density of SMBH seeds to be
only $n_{\rm SMBH} \sim 10^{-4}\:{\rm Mpc}^{-3}$. Wise et al. (2019)
carried out radiation-HD simulations and estimated a global co-moving
number density of DCBHs to be much smaller, i.e., at a level of
$\sim10^{-7}-10^{-6}\:{\rm Mpc}^{-3}$. The study of Latif et
al. (2022) of turbulence-supported halos also yields similar
constraints of $n_{\rm SMBH} < 10^{-6}\:{\rm Mpc}^{-3}$. For formation
from dense star clusters, it is difficult to calculate the occurence
rate of the conditions for efficient runaway growth. However the
densities needed in the models of G\"urkan et al. (2004) and Freitag
et al. (2006) are almost never seen in local populations of young
clusters (e.g., Tan et al. 2014).

One theory that aims to address the above features of the SMBH
population and form it with a single mechanism is SMBH birth from
Population III.1 stars (Banik, Tan \& Monaco 2019 [BTM19]; Singh,
Monaco \& Tan 2023 [SMT23]; Cammelli et al. 2024 [CMT24]). Here we
review the key features of the Pop III.1 model.

\begin{figure}[t]
\begin{center}
\includegraphics[width=4in]{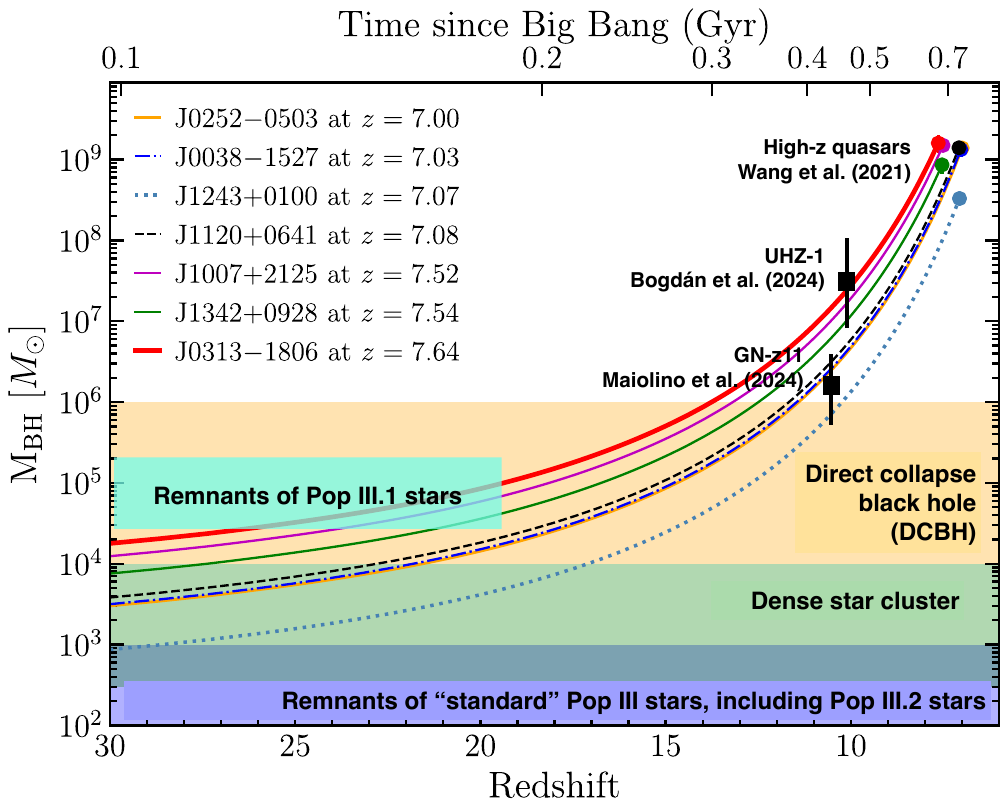}
\end{center}
\caption{
(adapted from Wang et al. 2021) Black hole growth tracks of a sample
of $z > 7$ quasars (circles) with the assumption of continuous
Eddington-limited accretion, i.e., with quasar luminosity
$L=\epsilon\dot{m}c^2=L_{\rm Edd}$ and a radiative efficiency of
$\epsilon = 0.1$. The curves give the mass of seed black holes
required to grow to the observed mass of each SMBH in these
quasars. With the above assumptions, these SMBHs require supermassive
($\sim10^5\:M_\odot$) seeds if the formation redshifts are $\sim
15-20$. The AGN at $z=10.1$ (UHZ-1, Bogdan et al. 2024) and $z=10.6$
(GN-z11, Mailoino et al. 2024) are also shown (black squares), with
similar implications. Shaded regions show mass scales of ``standard''
Pop III stellar remmnants (purple), IMBHs that may form in dense
stellar clusters (green), direct collapse black holes (yellow), and
Pop III.1 remnants (light blue) (see text).
}
\label{fig:1}
\end{figure}

\section{Pop III.1 Protostars - Impact of Dark Matter Annihilation}

Pop III.1 protostars are the first dense structures to form in their
local region of the Universe, situated at the center of ``isolated''
(i.e., undisturbed by radiative or chemical feedback from other
astrophysical sources) dark matter minihalos, i.e., with total masses
of $\sim10^6\:M_\odot$.  Under these conditions collapse of metal-free
gas to high densities occurs co-located with the peak of dark matter
density. Previous work on ``standard'' Pop III star formation in such
minihalos reached a near consensus view that the masses of the stars
formed are $\sim100-10^3\:M_\odot$ set by internal ionizing feedback
(e.g., Abel et al. 2002; Bromm et al. 2002; Tan \& McKee 2004;
McKee \& Tan 2008; Tan et al. 2010; Hosokawa et al. 2011; Hirano et
al. 2014; Susa et al. 2014; Tanaka et al. 2017). Such stars would
produce, at best, ``light'' seeds, i.e., $\lesssim 10^3\:M_\odot$ that
would have difficulties in explaining the observed high-$z$ AGN.

However, as first pointed out by Spolyar, Freese \& Gondolo (2008),
weakly interacting massive particle (WIMP) dark matter annihilation
(DMA) heating could have a major impact on Pop III star formation. In
models in which the WIMP is its own anti-particle and self-annihilates
with a rate coefficient given by $\langle \sigma_a v \rangle \simeq
3 \times 10^{-26}\:{\rm cm^3\:s}^{-1}$ (e.g., Jungman et al. 1996),
i.e., to explain the cosmic dark matter density, $\Omega_{\rm DM} =
0.268$, then a large fraction, $\sim 2/3$, of the energy, i.e., the
non-neutrino component consisting of high energy photons and
electron-positron pairs, is expected to be trapped in the star. The
rate of energy injection is boosted by many orders of magnitude if the
local dark matter density is enhanced by adiabatic contraction during
slow, monolithic contraction of the H/He gas in the minihalo.

Natarajan, Tan \& O’Shea (2009) examined several example simulated
minihalos confirming the results of Spolyar et al. (2008) that DMA
heating would become important in Pop III.1 protostars, including
setting their initial structures.
Smith et al. (2012) and Stacey et al. (2014) carried out simulations
of the first stages of Pop III.1 star formation including the effects
of WIMP annihilation on the chemistry and thermodynamics of the
collapse. These simulations found that by including WIMP DMA heating
there was much reduced fragmentation during the collapse, supporting
the likelihood that single massive star formation results from each
Pop III.1 minihalo. In addition, the amplification of magnetic fields
in Pop III accretion disks (Tan \& Blackman 2004; Latif \& Schleicher
2016) is also expected to inhibit fragmentation during the star
formation process. However, the final outcome of Pop III.1 star
formation from the simulations of Smith et al. and Stacey et
al. remains uncertain, as they were only able to follow the growth of
the star up to $\sim10\:M_\odot$.

\begin{figure}[t]
\begin{center}
\includegraphics[width=4in]{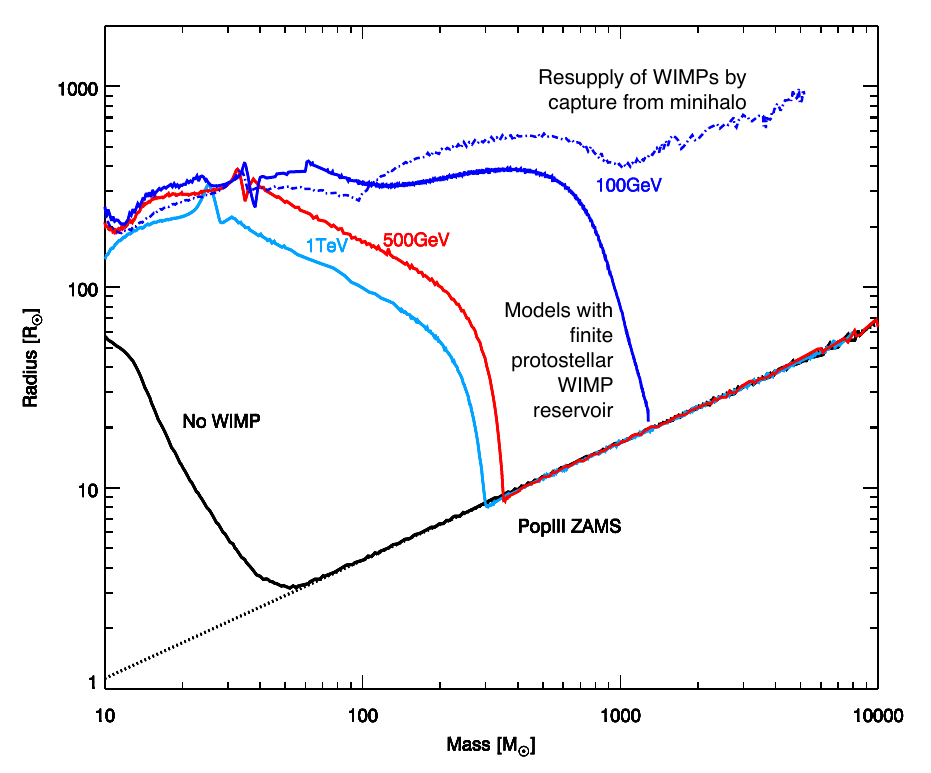}
\end{center}
\caption{
MESA (Modules for Experiments in Stellar Astrophysics; Paxton et
al. 2013) calculations of Pop III.1 protostars forming from a
minihalo with accretion rate given by the isentropic accretion model
of Tan \& McKee (2004). The standard (``no WIMP'') case is shown by
the black line, illustrating contraction to the zero age main sequence
(ZAMS) structure by a mass of $\sim 50\:M_\odot$. Accretion to higher
masses along the ZAMS is expected to halt at $\sim 100\:M_\odot$
because of photoevaporation feedback (e.g., McKee \& Tan 2008; Tanaka
et al. 2017). The light blue, red and dark blue solid lines show
models in which the protostar inherits a fixed amount of WIMP dark
matter based on its initial structure from the models of Natarajan et
al. (2009). WIMP annihilation heating keeps the protostar in a swollen
state until depletion of the initial WIMP reservoir by masses of
$\sim300$ to $1,000\:M_\odot$. The dashed blue line shows a model that
includes resupply of WIMPs by capture from the surrounding
minihalo. Here, the evolution has been followed to
$\sim5,000\:M_\odot$ with the star remaining in a large, swollen state
up to this point. Note, the main effect of WIMP annihilation heating
on protostellar structure is to keep the protostar large, cool, and
with a low ionizing feedback, thus potentially allowing accretion to
supermassive scales.
}
\label{fig:protostar}
\end{figure}

Freese et al. (2010) and Rindler-Daller et al. (2015) presented models
of protostellar evolution of DMA-powered protostars, which they refer
to as ``dark stars'', that continue to accrete to much higher
masses. In the study of Rindler-Daller et al. (2015), starting with
initial masses from 2 to $5\:M_\odot$, the protostars were followed to
$>10^5\:M_\odot$ for cases with WIMP masses of $m_\chi = 10, 100$ \&
$1000\:$GeV and gas accretion rates of $10^{-3}$ and
$10^{-1}\:M_\odot\:{\rm yr}^{-1}$. Typical accretion rates in Pop
III.1 minihalos are expected to be $\sim10^{-2}\:M_\odot\:{\rm yr}^{-1}$
(Tan \& McKee 2004), for which protostellar growth to
$\sim10^5\:M_\odot$ would take $\sim10^7\:$yr. The main feature of the
protostellar evolution calculations of Rindler-Daller et al. is that
the radius of the forming star remains very large (i.e., few $\times
10^3\:R_\odot$) and the photospheric temperature relatively cool
(i.e., $\sim 10^4\:$K). This implies that ionizing feedback from the
protostar will be relatively weak compared to the case of standard Pop
III star formation.
However, in these models Pop III.1 protostellar temperatures start to
rise above $10^4\:$K as $m_*$ grows beyond $10^3\:M_\odot$, so
internal feedback processes do need to be considered. In particular,
it remains to be demonstrated if the mass loss rate driven by internal
photoevaporation of the disk remains small compared to the accretion
rate. The photoevaporative mass loss rate is given by (McKee \& Tan
2008):
\begin{equation}
\dot{m}_{\rm evap} = 1.1 \times 10^{-4}S_{49}T_{\rm 3e4}^{0.4} m_{*d,3}^{1/2}\:M_\odot\:{\rm yr}^{-1},
\end{equation}
where $S_{49}\equiv S/10^{49}\:{\rm s}^{-1}$ is the protostellar
H-ionizing (EUV) photon production rate, $T_{\rm 3e4}\equiv
T/(3\times10^4\:{\rm K})$ is the ionized gas temperature (with
normalization appropriate for metal-free gas) and $m_{*d,3}\equiv
m_{*d}/(10^3\:M_\odot)$ is the mass of the star-disk system.

Another limitation of the Rindler-Daller et al. (2015) models is that
they do not allow for depletion of WIMPs within the protostar: i.e.,
they make the assumption of ``continuous replenishment'' by capture of
WIMPs from the surrounding minihalo. Figure~\ref{fig:protostar} shows
example Pop III.1 protostar calculations that relax this
assumption. Several models with finite WIMP reservoirs, implied by the
calculations of Natarajan et al. (2009) for various WIMP masses, are
shown. In these cases, contraction to the ZAMS is delayed until the
protostar has grown to at most $\sim 10^3\:M_\odot$. At this point
ionizing feedback would be expected to shut off growth of the
star. However, a model in which the protostar is able to keep
capturing WIMPs from the surrounding minihalo, also shown in
Fig.~\ref{fig:protostar}, allows the protostar to remain large and
relatively cool until at least $5,000\:M_\odot$. Thus, as discussed by
BTM19, it remains a plausible scenario that WIMP annihilation heating
could keep the protostar large, cool, and with relatively
low ionizing feedback that allows efficient accretion of the
baryonic content of the minihalo, i.e., $\sim10^5\:M_\odot$, to a
central, supermassive star, which, after a brief lifetime, ultimately
collapses to a SMBH of similar mass.

There are two main possibilities for how SMBH formation could occur in
these Pop III.1 minihalos. The protostar may run out of WIMP support,
i.e., fail to keep capturing enough WIMPs to maintain its luminosity,
leading to eventual contraction to the zero age main sequence
(ZAMS). At this point its ionizing feedback should shut off further
accretion. It would then undergo main sequence and post main sequence
stellar evolution, with the final outcome expected to be collapse of
most of the stellar mass to a black hole (e.g., Heger \& Woosley
2002). The other possibility is that the protostar keeps
accreting. Then collapse to a SMBH may be induced by the star becoming
unstable with respect to the general relativistic radial instability
(GRRI). For non-rotating main-sequence stars, this is expected to
occur at a mass of $\sim5 \times 10^4\:M_\odot$ (Chandrasekhar 1964),
while in the case of maximal uniform rotation this is raised to
$\sim10^6\:M_\odot$ (Baumgarte \& Shapiro 1999).

There are a few cases of SMBHs with masses below $10^5\:M_\odot$. For
example, H\"aberle et al. (2024) have estimated the mass of the black
hole in $\omega$Cen, a stripped nucleus of a dwarf elliptical galaxy,
to be in the range $0.8-5\times 10^4\:M_\odot$. Another example is the
dwarf galaxy RGG 118, which is estimated to host a
$\sim5\times10^4\:M_\odot$ nuclear BH (Baldassare et al. 2015). The
results of Mummery \& van Velzen (2024) imply the existence of a small
number of SMBHs just below $\sim 10^5\:M_\odot$. Pop III.1 protostars
are likely to have some rotational support and thus would not reach
the GRRI limit until $>10^5\:M_\odot$. Thus, we consider that the most
likely scenario involves Pop III.1 protostars running out of WIMP
support at mass scales $\gtrsim 10^4\:M_\odot$ and undergoing a phase
of main sequence and post main sequence nuclear burning before a core
collapse supernova explosion leads to efficent SMBH formation.



\section{Pop III.1, Pop III.2 and Cosmic Number Density of SMBHs}

\begin{figure}[t]
\begin{center}
\includegraphics[width=3.5in]{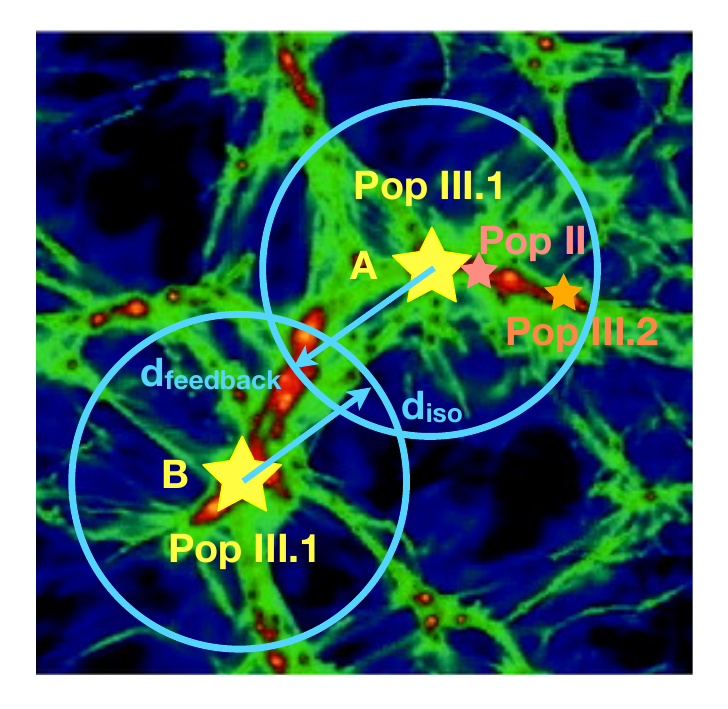}
\end{center}
\caption{
Schematic diagram illustrating the definition of Pop III.1 minihalos,
which are assumed to be SMBH progenitors, and Pop III.2 minihalos,
which form stars of $\sim10\:M_\odot$. Source A is the first to form
in this local region of the Universe. It irradiates its surroundings
out to a feedback distance, $d_{\rm feedback}$, which prevents other
Pop III.1 sources from forming. Instead, minihalos within this
irradiated volume form Pop III.2 stars if they have not been poluted
by stellar ejecta or Pop II stars if they have been. Source B forms
next as another Pop III.1 source, i.e., being alone within an
``isolation distance'' $d_{\rm iso}$, which is set equal to $d_{\rm
feedback}$.
}
\label{fig:schematic}
\end{figure}

The main parameter of the Pop III.1 theory is the ``isolation
distance'', $d_{\rm iso}$, that Pop III.1 halos need to be separated
from other astrophysical sources of radiation (see
Fig.~\ref{fig:schematic}). If a minihalo is too close to a source of
EUV radiation, i.e., closer than $d_{\rm iso}$, then its free electron
abundance is enhanced leading to greater rates of $\rm H_2$ formation,
enhanced cooling rates and higher degrees of fragmentation to
lower-mass ($\sim10\:M_\odot$) Pop III.2 stars (Greif \& Bromm 2006;
Johnson \& Bromm 2006). Fragmentation would imply that slow,
monolithic contraction of gas in the minihalo does not occur so that
dark matter densities are not enhanced via adiabatic contraction so
that there is a negligible impact of DMA heating on protostellar
evolution. Thus Pop III.2 minihalos do not give rise to SMBH seeds.

One potential physical model for the isolation distance is the size of
the ionized region around a supermassive Pop III.1 star. As discussed
above, it is possible that such stars exist for several Myr in a phase
on or near the main sequence, with photospheric temperatures of $\sim
10^5\:$K. The peak of the spectral energy distribution of such a
photosphere is at $\lambda_{\rm peak}\simeq 0.029\:{\rm \mu m}$,
corresponding to photon energies of $h c / \lambda = 43\:$eV.
The luminosity of such a star is expected to be near that of the
Eddington luminosity, i.e., $L_*\simeq L_{\rm Edd}= 3.2\times 10^9
(m_*/10^5\:M_\odot)\:L_\odot$. Most of the photons emitted by the star
have energies $>13.6\:$eV and so are able to ionize H and the rate of
production of these photons is $S\simeq 10^{53}\:{\rm s}^{-1}$.
Such a source would have an enormous impact on the surrounding
intergalactic medium (IGM). The extent of the ionized region can be
evaluated as the Str\"omgren radius:
\begin{eqnarray}
R_S & = & \left(\frac{3}{4\pi}\frac{S}{\alpha^{(2)}n_en_p}\right)^{1/3}
 = 61.3 S_{53}^{1/3}T_{\rm 3e4}^{0.27}\left(\frac{n_{\rm H}}{n_{\rm H,z=30}}\right)^{-2/3}\:{\rm kpc},\\
 & = & 1.90 S_{53}^{1/3}T_{\rm 3e4}^{0.27}(n_{\rm H}/n_{\rm H,z=30})^{-2/3} ([1+z_{\rm form}]/31)\:{\rm cMpc}\\
  & = & 1.90 S_{53}^{1/3}T_{\rm 3e4}^{0.27}([1+z_{\rm form}]/31)^{-1}\:{\rm cMpc}
\label{eq:hii}
\end{eqnarray}
where
$\alpha^{(2)}=1.08\times10^{-13}T_{\rm 3e4}^{-0.8}\:{\rm cm^3\:s^{-1}}$
is the recombination rate to excited states of ionized H,
$S_{53} \equiv S/10^{53}\:{\rm s}^{-1}$, $n_{\rm H}$ is number density
of H nuclei and $n_{\rm H,z=30}=5.72\times10^{-3}\:{\rm cm}^{-3}$ is
the average value of $n_{\rm H}$ in the IGM at a redshift of 30 (i.e.,
when most Pop III.1 sources are forming in the fiducial BTM19
scenario; see below). Note, for simplicity, we have ignored
reionization of He.

However, the time to establish ionization equilibrium in these HII
regions is relatively long, i.e., $t_{\rm ion}\simeq (4/3)\pi R_S^3
n_{\rm H} / S = (\alpha^{(2)}n_{\rm H})^{-1} = 51.3 [(1+z_{\rm
form})/31]^{-3} \:$Myr. This is much longer than the expected lifetime
of a supermassive star, i.e., $\lesssim 3\:$Myr, although residual DMA
heating would prolong its life somewhat. Taking a fiducial lifetime of
$t_*=10\:$Myr, we estimate the extent of an R-type expanding HII
region:
\begin{eqnarray}
R_R & = & \left(\frac{3 t_* S}{4\pi n_{\rm H}}\right)^{1/3} = 35.5 t_{*,10}^{1/3} S_{53}^{1/3} (n_{\rm H}/n_{\rm H,z=30})^{-1/3}\:{\rm kpc}\\
 & = & 1.10 t_{*,10}^{1/3} S_{53}^{1/3} (n_{\rm H}/n_{\rm H,z=30})^{-1/3} (1+z_{\rm form})/31\:{\rm cMpc}\\
  & = & 1.10 t_{*,10}^{1/3} S_{53}^{1/3}
\:{\rm cMpc},
\end{eqnarray}
where $t_{*,10} = t_*/10\:$Myr. Note, that the co-moving extent of an
R-type HII region is independent of redshift. The above estimate for
the radius of the HII region undergoing R-type expansion assumes fully
ionized conditions inside $R_R$. To check this approximation, we
evaluate the mean free path for ionizing photons as
\begin{equation}
\lambda_{\rm mfp} = (n_H \sigma_{\rm p.i.})^{-1} = 9.0 (n_{\rm H}/n_{\rm H,z=30})^{-1} (h\nu/13.6\:{\rm eV})^{3} \:{\rm pc}.
\end{equation}
Thus even for very energetic photons, i.e., with $h\nu \sim 100\:$keV, then $\lambda_{\rm mfp}$ approaches $3.6\:$kpc, which is still small compared to $R_R$.

Assuming feedback distance $d_{\rm feedback}=R_R$ and that the
isolation distance for Pop III.1 SMBH formation is set by this
feedback, i.e., $d_{\rm iso}=d_{\rm feedback}$, we estimate a
co-moving number density of SMBHs as
\begin{equation}
n_{\rm SMBH}= \frac{3}{4\pi R_R^3} \rightarrow  0.18 t_{*,10}^{-1} S_{53}^{-1}\:{\rm cMpc}^{-3}.\label{eq:nlim}
\end{equation}
This estimate can be viewed as an upper limit since it has assumed
maximal close packing of the sources and ignored other contributions
to the ionizing background from Pop III.2, Pop II and AGN sources.

One of the main predictions of the Pop III.1 model for SMBH formation
is that all SMBHs form very early in the Universe and are born in
locally isolated minihalos, i.e., as single SMBHs. Thus their initial
distribution is relatively unclustered. Clustering and mergers of
seeded halos would occur much later as large-scale structure
develops. Another main prediction is the existence of large ($\sim
1\:$cMpc) HII regions around these first sources. These HII regions
would have a large volume filling factor at high redshift, but would
then recombine after a relatively short time, i.e., $\sim
50\:$Myr. Nevertheless, the residual free electrons in these relic HII
regions are reason for suppressing further SMBH formation, since they
lead to enhanced $\rm H_2$ formation in the affected Pop III.2
minihalos, resulting in fragmentation to $\sim 10\:M_\odot$ stars. In
the next section we review the results of implementation of the Pop
III.1 seeding scheme in simulations of cosmological volumes.

\begin{figure}[t]
\begin{center}
\includegraphics[width=3.5in]{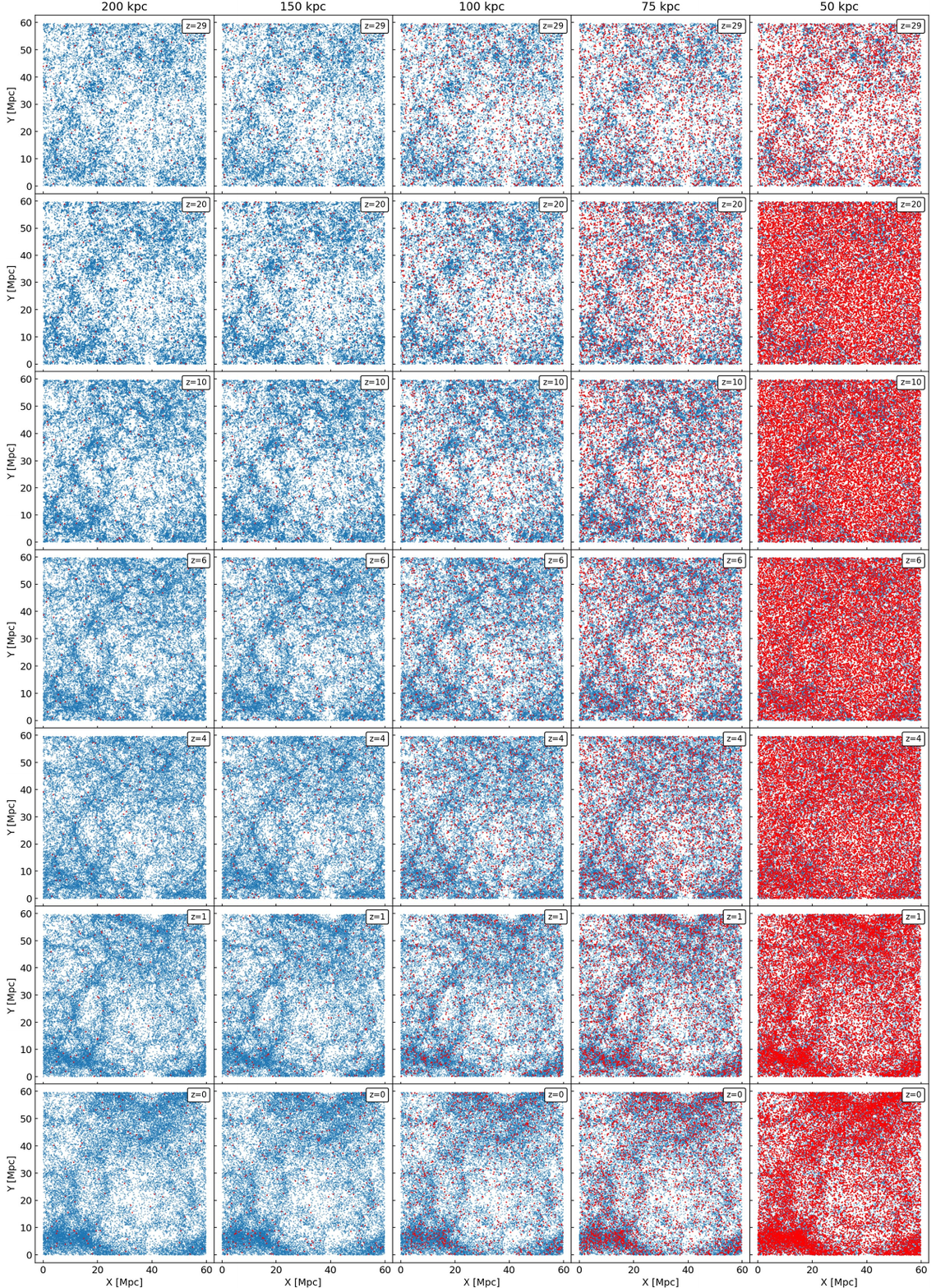}
\end{center}
\caption{
(from SMT23) Projection of the positions of seeded haloes (red) and
non-seeded haloes (blue) in the XY plane of a $(\sim60\:{\rm cMpc})^3$
volume for different isolation distances. The redshift is shown in the
top right corner of each panel. Only the 30,000 most massive
non-seeded halos within each panel are shown for ease of
visualization.
}
\label{fig:sim}
\end{figure}

\section{Cosmological Simulations of Pop III.1 Seeding of SMBHs and Comparison to Observational Constraints}

BTM19 presented simulations of a $\sim (60\:{\rm cMpc})^3$ volume
using the PINOCCHIO (PINpointing Orbit Crossing Collapsed HIerarchical
Objects) code (Monaco et al. 2002; Monaco et al. 2013), which uses a
Lagrangian Perturbation Theory (LPT) based method (Moutarde et
al. 1991; Buchert \& Ehlers 1993; Catelan 1995) for the fast
generation of catalogues of dark matter haloes in cosmological
volumes. The BTM19 study had a dark matter particle mass resolution of
$\sim10^5\:M_\odot$, so that minihalos were resolved at a level of 10
particles per halo, sufficient to identify their location, and the
evolution was followed down to $z = 10$. The isolation distance
parameter, $d_{\rm iso}$, was set equal to a constant value of proper
distance, with values of $d_{\rm iso}$ from 10 to 300 kpc
explored. SMT23 extended this type of calculation down to $z =
0$. Figure~\ref{fig:sim} shows example outputs, i.e., projections of
seeded and unseeded halos across the simulation volumes from high
to low redshifts.
SMT23 also explored an alternative seeding scheme based on a simple
halo mass threshold (HMT), i.e., halos are seeded with a SMBH once
they exceed a mass of $7.1 \times 10^{10}\: M_\odot$, which is a
method commonly used in cosmological volume galaxy formation
simulations, e.g., Illustris-TNG (Vogelsberger et al. 2014; Weinberger
et al. 2017).

To better connect the predictions of the Pop III.1 SMBH seeding to
observable galaxy properties one needs to implement a model of galaxy
formation and evolution in the simulations. To this end, CMT24 applied
the GAlaxy Evolution and Assembly (GAEA) semi-analytic model (Fontanot
et al. 2020) to the merger histories of the PINOCCHIO simulated
halos. CMT24 considered Pop III.1 seeding for $d_{\rm iso}=50, 75$ and
100~kpc. They also compared these models with the HMT seeding model
(with threshold halo mass of $7.1\times10^{10}\:M_\odot$) and with an
All Light Seeds (ALS) seeding scheme, which seeds all halos with
low-mass black holes.

\subsection{Co-Moving Number Density of SMBHs}

\begin{figure}[t]
\begin{center}
\includegraphics[width=3.75in]{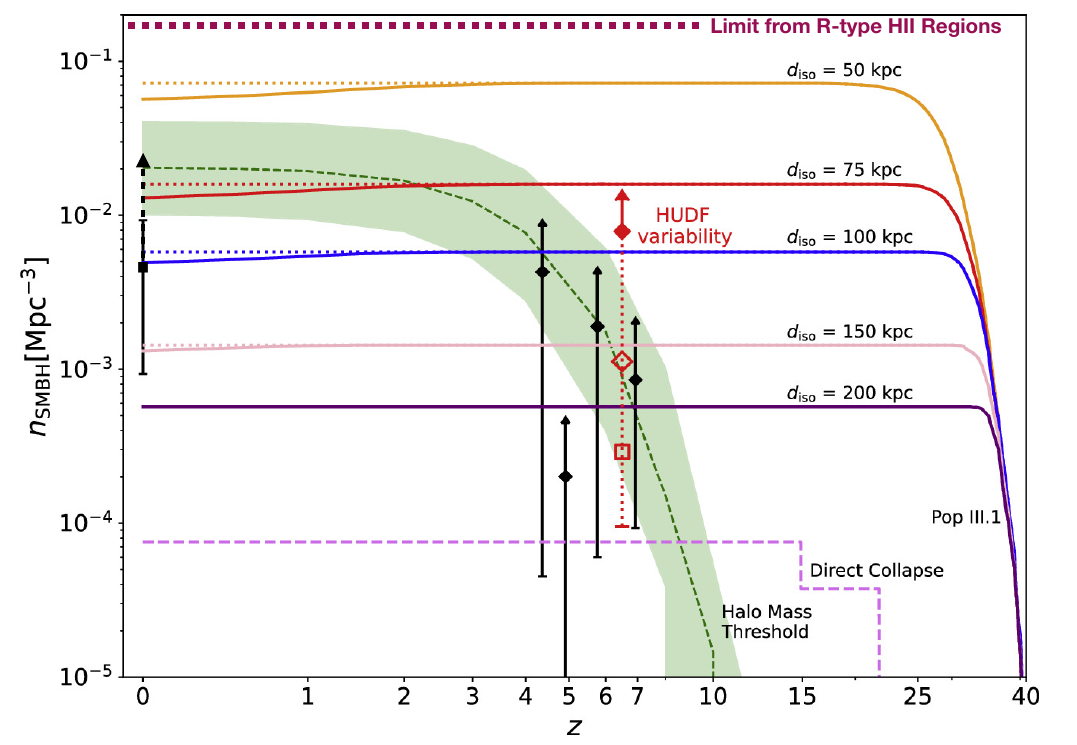}
\end{center}
\caption{
Cosmic evolution of SMBH abundance. The comoving number density of
SMBHs, $n_{\rm SMBH}$, is plotted vs redshift, $z$. Various Pop III.1
formation models (BMT19; SMT23) with isolation distance parameters,
$d_{\rm iso}$, from 50 to 200 kpc (proper distance) are shown by the
colored solid lines. At low redshifts, these decrease compared to the
maximum value attained (dotted lines) due to mergers. The dark red
dotted line at $0.18\:{Mpc}^{-3}$ is the close-packed limiting value
from the fiducial R-type HII region expansion model for $d_{\rm iso}$
(eq.~\ref{eq:nlim}).
The green dashed line shows SMBH seeding based on an HMT above a mass
of $7.1 \times 10^{10}\:M_\odot$ (Vogelsberger et al. 2014; Weinberger
et al. 2017), with the shaded region illustrating a factor of 2
variation in this mass scale. The magenta dashed line shows results
from a direct collapse model of SMBH formation (Chon et al. 2016). An
observational constraint derived by counting $z = 6–7$ AGN in the HUDF
variability survey (Hayes et al. 2024) is shown by the red open
square, and the lower bound shows the Poisson uncertainty. The red
open diamond corrects for variability incompleteness, and the red
filled diamond is for luminosity incompleteness (corrected down to
$M_{\rm uv}=-17\:$mag). Observational lower limits at $z = 0$ from
local galaxies (BTM19) and at $z = 4–7$ from broad emission line AGN
(Harikane et al. 2023) are shown with black square and diamonds,
respectively.
%
}
\label{fig:nevol}
\end{figure}

Figure~\ref{fig:nevol} shows the redshift evolution of the co-moving
number density of SMBHs, $n_{\rm SMBH}$, for Pop III.1 models with
$d_{\rm iso}$ from 50 to 200~kpc. Results of SMBH seeding based on an
HMT above a mass of $7.1 \times 10^{10}\:M_\odot$ are also shown, as
well as the Direct Collapse simulation of Chon et al. (2016). We see
that the Pop III.1 models predict the existence of high number
densities of SMBHs (and thus also AGN) at high $z$. A value of $d_{\rm
iso}=100\:$kpc is sufficient to produce $n_{\rm SMBH}\simeq 5\times
10^{-3}\:{\rm Mpc}^{-3}$, which is inferred from local SMBH
populations (BTM19). However, it should be noted that, due to
incompletenees, this is likely to be a lower limit. Thus smaller
values of $d_{\rm iso}$, i.e., closer to $50\:$kpc, may be needed. For
values of $d_{\rm iso}\gtrsim 50\:$kpc, we see from
Fig.~\ref{fig:nevol} that essentially all SMBHs are in place by
$z=20$. After this $n_{\rm SMBH}$ remains nearly constant down to low
redshifts, with only modest (up to $\sim 20\%$) decreases due to
mergers that mainly occur at $z\lesssim2$.

One of the main discoveries from early JWST observations has been the
large numbers of AGN found at high redshifts. Fig.~\ref{fig:nevol}
shows estimates at $z=4 - 7$ presented by Harikane et al. (2023), from
the sample of Nakajima et al. (2023), with AGN identified by the
presence of broad emission lines. These estimates, which involve an
incompleteness correction down to a magnitude limit of $M_{\rm
UV}=-17\:$mag, are lower bounds on $n_{\rm SMBH}$, since fainter AGN
are also expected to be present.

Hayes et al. (2024) carried out a re-observation of the Hubble Ultra
Deep Field (HUDF) in 2023 using the Hubble Space Telescope (HST) with
WFC3/IR in the F140W filter. This was designed to match the 2012
observation of Ellis et al. (2013) in the same filter in order to
search for AGN via photometric variability. In addition, observations
in 2009 (Bouwens et al. 2010) were compared with those of Ellis et
al. in filters F105W and F160W. From a first examination of these
data, three AGN were identified with high significance at $z=6-7$. A
variability incompleteness factor of $\sim$5 is estimated (Cammelli et
al., in prep.), given the recovery rate of 31 known AGN in the HUDF
(Lyu et al. 2022). Luminosity incompleteness correction down to
$M_{\rm UV}=-17\:$mag implies a factor of $\sim$7 boost, yielding
$n_{\rm SMBH}\gtrsim 8\times 10^{-2}\:{\rm cMpc^{-3}}$. From these
results (and those of the more comprehensive analysis of the full
dataset; Cammelli et al., in prep.) we infer $d_{\rm iso}\lesssim
75\:$kpc. The measurements of $n_{\rm SMBH}$ at high $z$ are
strikingly similar to estimates in the local Universe and these begin
to place meaningful constraints on both Pop III.1 and HMT type models.

\subsection{Clustering of SMBHs}


Since SMBHs are born isolated from other sources, i.e., by a distance
$d_{\rm iso}$, which is $\sim 1-3\:$cMpc, then SMBHs are predicted to
have a low degree of initial spatial clustering. BTM19 and SMT23 found
much lower levels of clustering of seeded halos, e.g., as measured via
two-point angular or spatial correlation functions, compared to
equivalently massive halos selected independently of SMBH seeding. The
level of SMBH clustering increased towards lower redshifts as
large-scale structure developed, but the imprint of $d_{\rm iso}$
remained visible down to $z\sim 1$. Note, that the Pop III.1 model
nevertheless predicts that SMBHs are found in highly biased regions,
since these are the first-forming halos in each local region of the
Universe. Comparisons of these results with the clustering properties
of AGN (e.g., Hennawi et al. 2006; Hickox et al. 2009) will provide
further tests of the Pop III.1 models.

\subsection{Multiplicity of SMBHs}

In the Pop III.1 scenario all SMBHs are born as single objects and,
being isolated from each other, the level of binary and higher-order
multiplcity, i.e., 2 or more SMBHs in the same halo, is very low at
high $z$. Precise estimates of the binary SMBH fraction depend
sensitively on the assumed timescales for galaxies to merge once halos
have merged and for the lifetime of the SMBH pair within the same
galaxy before their merger. However, from preliminary estimates (Singh
2024; Singh et al., in prep.) it is found that for $d_{\rm iso}\gtrsim
75\:$kpc there are essentially no binary AGN at $z>5$. For $d_{\rm
iso}=50\:$kpc and an assumed halo to galaxy merger time of 500~Myr and
a pair lifetime of 500~Myr, the binary SMBH fraction at $z\sim 5$ is
$\sim 10^{-4}$, with the fraction rising to few~$\times 10^{-2}$ by
$z=2$.

Yue et al. (2021) reported a close ($\lesssim 10\:$kpc-scale) quasar
pair at $z=5.66$. They used this object to infer a lower limit on the
SMBH pair fraction at $z=5-6$ on scales $<30\:$kpc of $f_{\rm
pair}>3\times 10^{-3}$. Yue et al. (2023) reported discovery of a
kpc-scale quasar pair at $z=4.85$. Maiolino et al. (2024b) presented
three high-$z$ ($z = 4.1, 4.4, 5.9$) candidate dual AGN, but with
multiplicity inferred only spectroscopically. Matsuoka et al. (2024)
reported a quasar pair separated by 12 kpc in projected distance at
$z=6.05$. From these results, we infer that $d_{\rm iso}\lesssim
50\:$kpc is preferred. However, further work is needed to investigate
the sensitivity of the Pop III.1 simulation results to assumed galaxy
and SMBH merger timescales.

\subsection{SMBHs and Galaxy Evolution}

\begin{figure}[t]
\begin{center}
\includegraphics[width=4.0in]{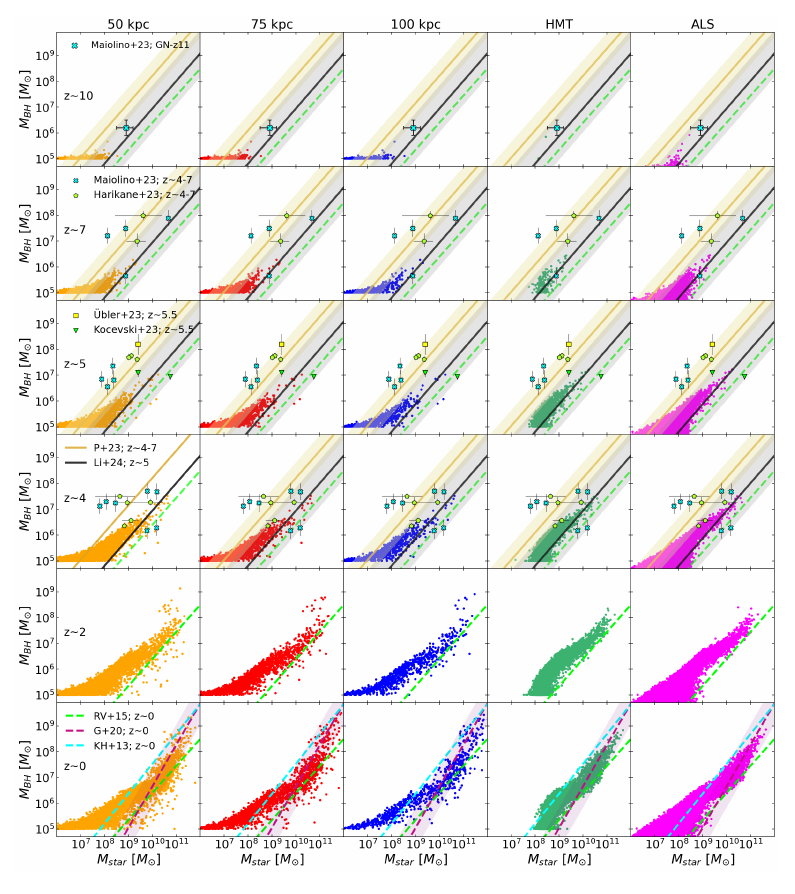}
\end{center}
\caption{
(From CMT24) Scatter plot of the evolution of the $M_{\rm BH}-M_{\rm
star}$ relation at several redshifts (rows, as labelled). First three
columns are Pop III.1 seeding schemes with $d_{\rm iso}=50, 75,
100\:$kpc. Fourth column is the fiducial Halo Mass Threshold (HMT)
scheme and fifth column is the All Light Seed (ALS) scheme (see text).
At redshift $z\sim 0$ several local empirical relations are shown:
Kormendy \& Ho (2013) (KH+13); Reines \& Volonteri (2015) (RV+15);
Greene et al. (2020) (G+20). At higher $z$ the RV+15 $z=0$ fit is
shown for reference. High-$z$ rows show comparison with JWST-detected
AGN: in Pacucci et al. (2023), they directly fit the data in the
redshift range $z\sim4-7$, while Li et al. (2024) estimate an unbiased
fit taking into account the uncertainties on the mass measurements and
selection effects. Shaded regions illustrate the intrinsic scatter at
1-$\sigma$ according to each relation. Colored symbols show faint AGN
from Maiolino et al. (2024b); Harikane et al. (2023); \"Ubler et
al. (2023); Kocevski et al. (2023) and reported according to their
redshifts. In the top row, results at $z \sim 10$ are shown against
the single data point (GN-z11) from Maiolino et al. (2024).
%
}
\label{fig:mbh_mstar}
\end{figure}

A main predictions of the Pop III.1 model is that SMBHs form before
their eventual host galaxies. Thus the ratio of SMBH mass to host
galaxy stellar mass will tend to be relatively large in high-$z$
systems compared to those of the local Universe. However, detailed
predictions of the co-evolution of SMBHs and galaxies are sensitive to
the uncertain processes of star formation, SMBH fuelling, and
feedback.

CMT24 have presented models of SMBH and galaxy co-evolution for the
Pop III.1 seeding scenario based on the GAEA semi-analytic model
(Fontanot et al. 2020). Figure~\ref{fig:mbh_mstar} shows the evolution
of the $M_{\rm BH}-M_{\rm star}$ relation of these models versus
redshift, also comparing to an HMT model with threshold halo mass of
$7.1\times10^{10}\:M_\odot$ for SMBH seeding and an All Light Seeds
(ALS) model.

For the $M_{\rm BH}-M_{\rm star}$ relations at $z\sim0$, compared to
HMT and ALS, the Pop III.1 models tend to have: 1) larger dispersions;
2) shallower low-mass indices; 3) steeper high-mass indices; 4) a
larger number of the most massive SMBHs. As discussed by CMT24, these
differences are potentially testable via comparison with censuses of
local SMBH populations, including via comparison to SMBH mass
functions (e.g., Mutlu-Pakdil et al. 2016; Shankar et al. 2020;
Liepold \& Ma 2024).

At redshifts of $z\sim5$, the differences between the different
seeding models for the $M_{\rm BH}-M_{\rm star}$ relation are
relatively modest. The overall normaliztion and index of these
relations are consistent with the high-$z$ estimate of Li et
al. (2024), which, because of different analysis methods affecting
accounting for survey biases, is about 1 dex below the relation
inferred by Pacucci et al. (2023) from similar data.

Only at the highest redshifts, $z\gtrsim7$ do differences between
seeding schemes become more pronouced. For HMT this is because of the
lack of massive halos and thus SMBHs at these redshifts. For ALS the
black holes and galaxies grow together from very low masses, but by
$z\sim10$ there can be examples of $\sim 10^6\:M_\odot$ SMBHs that are
comparable to GN-z11. All models struggle to reach the $\sim
10^7-10^8\:M_\odot$ masses of the JWST samples. This may be caused by
the simulated box being quite small, i.e., $\sim(60\:{\rm Mpc})^3$,
and the observed AGN being quite rare objects. Or it may reflect
limitations in the semi-analytic SMBH growth model.

To investigate this issue further, using the RAMSES code (Rosdahl et
al. 2013), full radiation hydro simulations of SMBH growth and galaxy
formation around a Pop III.1 seed have been implemented (Sanati et
al., in prep.). A supermassive star phase with $S_{\rm 53}=1$ for
10~Myr is implemented in the first minihalo of a $(7\:{\rm Mpc})^3$
volume. We note that while this minihalo is identified at $z=29$, it
takes $\sim50\:$Myr, i.e., $z\sim 22$, for gas to reach high enough
densities for Pop III.1 star formation. Following the assumed 10~Myr
lifetime, when surrounding gas is mostly ionized, the source is then
converted into a sink particle of mass $10^{5}\:M_\odot$ that
represents the growing SMBH, including its thermal, kinetic and
radiative feedback. By $z=9$ the SMBH has reached $\sim 3\times
10^7\:M_\odot$, while the host galaxy has $M_{\rm
star}\sim10^8\:M_\odot$. Such properties are similar to the
JWST-detected systems at $z\gtrsim 7$.

Predictions for the luminosity functions of high-$z$ AGN and galaxies
of CMT24 have been carried out (Cammelli et al., in prep.). The
overall luminosity function, i.e., of both seeded and unseeded
galaxies, matches observed data well in the range $M_{\rm
UV}\gtrsim-22\:$mag out to $z\sim 10$. A main feature of the Pop III.1
models is the relatively high and constant amplitude of the AGN + host
galaxy luminosity function at $z\gtrsim7$, especially in comparison to
that of the HMT model. We note that larger volumes are needed to
sample rare objects (relevant to $M_{\rm UV}\lesssim-22\:$mag).

Finally, as discussed by Ilie et al. (2023), we note that emission
from Pop III.1 supermassive stars themselves may contribute to the
observed luminosity functions, especially at the highest
redshifts. Ilie et al. have modeled JWST-detected sources out to
$z\simeq 14$ with supermassive ($\sim10^6\:M_\odot$) ``dark star''
spectra. Detailed predictions of such emission in the context of the
Pop III.1 cosmological model are still needed. However, if $d_{\rm
iso}\gtrsim50\:$kpc, then we expect most such stars only at
$z\gtrsim20$.

\subsection{Occupation Fraction of SMBHs}

\begin{figure}[t]
\begin{center}
\includegraphics[width=4in]{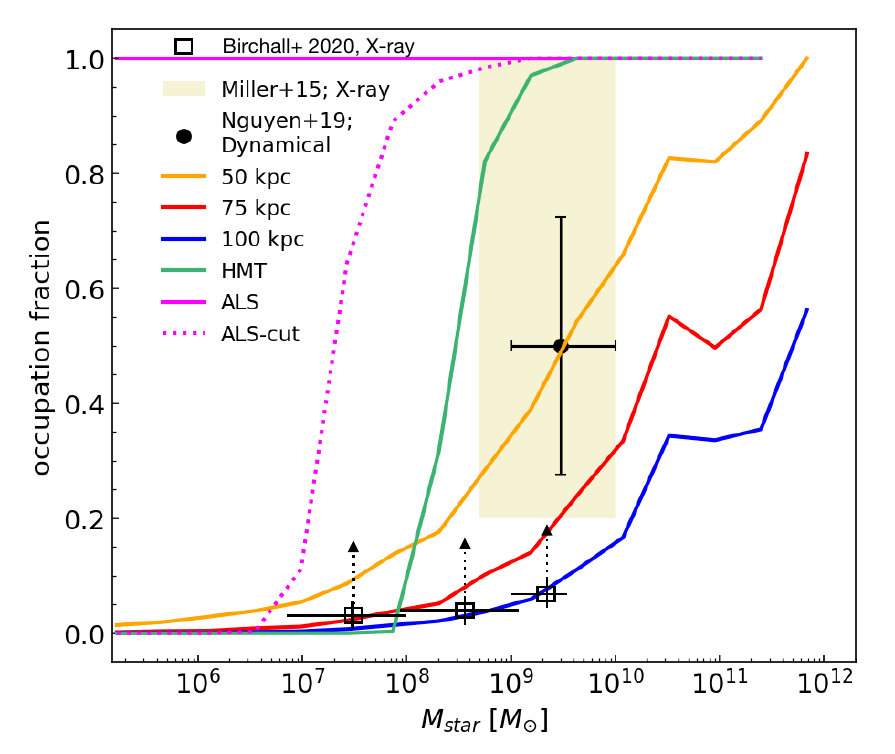}
\end{center}
\caption{
(Adapted from CMT24) Occupation fraction of seeded galaxies as a
function of stellar mass at redshift zero. The yellow, red and blue
lines show Pop III.1 models with $d_{\rm iso}=50, 75, 100\:$kpc. The
green line shows an HMT model with threshold halo mass of
$7.1\times10^{10}\:M_\odot$ for SMBH seeding. The solid pink line is
an All Light Seeds (ALS) model, which seeds all halos with low-mass
black holes, so has an occupation fraction of unity. The dotted pink
line shows the ALS results only for SMBHs with masses
$>10^5\:M_\odot$. The following observational constraints are shown:
an X-ray survey result that yields a lower limit of about 0.2 in
systems with $M_{\rm star}\sim 10^9-10^{10}\:M_\odot$ (Miller et
al. 2015); an X-ray survey of lower-mass systems that detects AGN (with
${\rm log_{10}} L_X / {\rm erg\:s}^{-1}>39$) in about 3-6\% of the
systems (Birchall et al. 2020); a study of the dynamical properties of
the nuclei of a small sample of nearby galaxies (Nguyen et al. 2019).
%
}
\label{fig:occupation}
\end{figure}

BTM19 and SMT23 analyzed the SMBH occupation fractions, $f_{\rm
SMBH}$, of dark matter halos, i.e., what fraction of halos host SMBHs
as a function of halo mass. For the case of $d_{\rm iso}=50\:$kpc, by
redshift of zero, $f_{\rm SMBH}$ reached unity for halos with masses
$\gtrsim10^{12}\:M_\odot$, while a level of $\sim 10\%$ seeding was
present down to halo masses of $\sim 10^{10}\:M_\odot$. As expected,
occupation fractions become smaller as $d_{\rm iso}$ increases.

The CMT24 results for $f_{\rm SMBH}$ as a function of galaxy stellar
mass are shown in Fig.~\ref{fig:occupation}. We see that a major
difference between the different $d_{\rm iso}$ models and the HMT and
ALS models is the occupation fraction for galaxies with $M_{\rm
star}\sim10^9\:M_\odot$. This figure also shows observational
constraints, which, however, have large uncertainties. Nevertheless,
these results indicate a preference for $d_{\rm iso}\lesssim 75\:$kpc.

\subsection{Gravitational Wave Emission of SMBHs}

Analysis of the North American Nanohertz Observatory for Gravitational
Waves (NANOGrav) pulsar timing array (PTA) 15 yr data set has detected
a signal consistent with that expected from a nHz gravitational wave
background (GWB) with a false-alarm probability of
$\sim10^{-4}-10^{-3}$ ($\simeq 3\sigma$) (Agazie et al. 2023a). The
most natural explanation for this signal is the result of the cosmic
population of SMBH binaries (e.g., Agazie et al. 2023b). The planned
Laser Interferometer Space Antenna (LISA) will operate at mHz
frequencies and be sensitive to mergers of SMBH binaries with masses
$\sim10^4-10^8\:M_\odot$ out to $z\sim 10$ (Amaro-Seoane et al. 2023).

The Pop III.1 model can be used to make predictions for the
gravitational wave emission that emerges from the cosmic population of
SMBH binaries. As mentioned above, one feature of Pop III.1 seeding is
that SMBHs are born relatively isolated from each other. Thus, as
shown in Fig.~\ref{fig:nevol}, mergers of seeded halos that would
produce SMBH binaries occur only at relatively low redshifts, i.e.,
$z\lesssim 2$. Naturally, mergers start earlier and are more frequent
in the models with smaller $d_{\rm iso}$. SMT23 find that by $z=0$
(and in the limit of fast merging of SMBHs once two seeded halos
merge) the raw number of mergers in a $\sim (60\:{\rm cMpc})^3$ volume
rises from 191 (out of a total population of 1,234, i.e., 15.5\%) for
$d_{\rm iso}=100\:$kpc to 3,305 (out of a total population of 15,400,
i.e., 21.5\%) for $d_{\rm iso}=50\:$kpc.

To make predictions for the GWB from a given SMBH seeding model, two
main astrophysical processes need to be modeled: 1) the growth of
SMBHs; 2) the timescale for SMBH binary merging following halo
merger. Each of these processes are complicated and it is difficult to
make accurate predictions (see, e.g., Agazie et al. 2023b). For SMBH
growth, as an alternative to modeling, one can resort to empirical
scaling relations between host galaxy properties and SMBH mass.

Preliminary work examining the implications of the Pop III.1 model for
the GWB has been presented by Singh (2024), but with more detailed
calculations underway (Singh et al., in prep.). Preliminary findings
are that the nHz GWB can be reproduced if reasonable, empirical
assumptions about SMBH growth are made, however, the results are
sensitive to treatment of the most massive SMBHs and their merger
rates. Given the relatively large contribution from the most massive
systems, the amplitude of the GWB is relatively insensitive to $d_{\rm
iso}$, i.e., with only a factor of $\sim3$ increase as $d_{\rm iso}$
decreases from 100~kpc to 50~kpc. We anticipate that greater
sensitivity to $d_{\rm iso}$ will be present in higher order
statistics of the GWB, e.g., the relative contribution of strongest
individual sources, which also impart large scale asymmetries in the
GWB (e.g., Gardiner et al. 2024).

In terms of predictions for the rate of SMBH merger events that will
be detectable by LISA, scaling to the volume of the Universe out to
$z=2$, i.e., $\sim600\:{\rm cGpc}^3$, from the results of SMT23 for
$d_{\rm iso}=100\:$kpc we expect $5.3\times10^8$ mergers to occur. In
the limit of fast merging time of SMBHs after halo merger and if these
occur at a uniform rate over the lookback time of $\sim 10\:$Gyr, then
this implies an event rate of $0.05\:{\rm yr}^{-1}$. The equivalent
rate for $d_{\rm iso}=50\:$kpc is $0.9\:{\rm yr}^{-1}$. Most of these
mergers are expected to be detectable by LISA, especially given a
fiducial Pop III.1 seed mass of $\sim 10^5\:M_\odot$. Thus we see that
the overall event rate that LISA may detect, e.g., in a nominal 4-year
mission lifetime, is very sensitive to the isolation distance
parameter. Another key prediction of the Pop III.1 models is the
redshift distribution of the events. For the $d_{\rm iso}=50\:$kpc
case this extends to moderately higher redshifts ($z\simeq2.5$)
compared to the $d_{\rm iso}=100\:$kpc case ($z\simeq1.5$). However,
the main prediction is that all events should be at these relatively
low redshifts.

\subsection{Contribution to Cosmic Reionization}

A key prediction of the Pop III.1 model is the existence of large
($\sim1\:$cMpc) HII regions around supermassive Pop III.1 stars at
$z\sim 20-30$. These HII regions fill a large fraction of the volume
of the Universe at these redshifts, but recombine on timescales of
$\sim 50 - 200\:$Myr, i.e., down to $z\sim 15$. Such sources could
make a significant contribution to the Thomson scattering optical
depth of CMB photons. From preliminary modeling using the BEoRN code
(Schaeffer et al. 2023) of such HII regions around Pop III.1 stars (la
Torre et al., in prep.), this contribution to the total $\tau_{\rm
CMB}\simeq 0.055$ rises from $\sim$ 6\% for $d_{\rm iso}=100\:$kpc to
$\sim36\%$ for $d_{\rm iso}=50\:$kpc.

The $E-$mode CMB polarization power spectrum has some sensitivity to
the redshift dependence of reionization history (e.g., Heinrich et
al. 2017; Millea \& Bouchet 2018). Future CMB polarization
observations, e.g., with LiteBIRD (Allys et al. 2023) will be able to
improve on such contraints. In addition, future CMB observations with
the Simons Observatory (Ade et al. 2019) will be able to test the Pop
III.1 prediction of the presence of an early phase of partial cosmic
reionization, in particular via observation of the ``patchy''
kinematic Sunyaev-Zel'dovich (kSZ) effect from the peculiar motion of
free electron bubbles around these first stars.

The ionized bubbles around Pop III.1 sources may also be detectable in
surveys of high-$z$ 21~cm emission. For example, the Hydrogen Epoch of
Reionization Array (HERA; Abdurashidova et al. 2022) has sensitivity
to such emision out to $z\sim 28$ and may be able to detect if there
is a characteristic size of such bubbles.

\section{Implications for Dark Matter}

The Pop III.1 mechanism for SMBH formation {\it requires} energy
injection from decaying dark matter. This dark matter particle has so
far been assumed to be a WIMP (e.g., Spolyar et al. 2008; Natarajan et
al. 2009; Rindler-Daller et al. 2015). While there is some dependence
of the protostellar initial condition and early evolution to the WIMP
mass (Natarajan et al. 2009; see Fig. 2), we anticipate that the
primary sensitivity of the model to particle properties is via the
spin-dependent scattering cross section, i.e., of WIMPs with protons.
This is because for growth to supermassive scales in a large, swollen
state the protostar needs to continue capture WIMPs from the
surrounding minihalo. However, to make accurate calculations of this
rate and its sensitivity to assumed particle properties (e.g., Aalbers
et al. 2024), requires accurate knowledge of the surrounding dark
matter density. This requires modeling adiabatic contraction of the
dark matter in the minihalo caused by the slow baryonic contraction of
the Pop III.1 protostellar infall envelope. Finally, given the current
non-detection of WIMPS, we note that more exotic forms of cold dark
matter, even including very low-mass primordial black holes (e.g.,
Carr \& Kuhnel 2020), could allow the Pop III.1 mechanism to operate,
with the main requirement that the dark matter inject energy inside
these early protostars.

\section{Summary}\label{sec:summary}

The Pop III.1 cosmological model for SMBH formation (BTM19, SMT23,
CMT24) invokes dark matter annihilation heating in first-forming,
undisturbed minihalos, where slow baryonic collapse leads to adiabatic
contraction of the dark matter density. Only under such specialized
conditions does the dark matter particle number density reach high
enough levels for there to be significant annihilation heating that
impacts protostellar structure (Spolyar et al. 2008; Natarjan et
al. 2009; Rindler-Daller et al. 2015) to maintain large, cool
photospheres that have limited ionizing feedback allowing growth of
supermassive Pop III stars and thence SMBHs.

The Pop III.1 model naturally produces a characteristic mass of SMBH
seeds of $\sim 10^5\:M_\odot$, with this related to the baryonic
content of $\sim 10^6\:M_\odot$ dark matter minihalos. In contrast to
other formation theories, the Pop III.1 model explains why there is a
paucity of IMBHs. A phase of ionizing feedback from supermassive Pop
III.1 protostars that emit $\sim 10^{53}$ H-ionizing photons per
second sets the overall cosmic abundance of SMBHs, which is expected
to be $\lesssim 0.2\:{\rm cMpc}^{-3}$. Most minihalos, having been
exposed to ionizing radiation that elevates their free electron
abundance, fragment into $\sim 10\:M_\odot$ Pop III.2 stars that do
not adiabatically contract their dark matter. The SMBH seed population
is formed very early: they are all in place by $z\sim 20$. After this
the SMBH co-moving number density is nearly constant until mergers
begin to occur by $z\lesssim 2$. Additional predictions are: a low
level of initial clustering of SMBHs; a low level of binary SMBHs (and
associated mergers) at high $z$; and ionized, $\sim1\:$cMpc-sized
bubbles at $z\gtrsim20$ that contribute to partial early cosmic
reionization.

To the extent that the Pop III.1 model is needed to explain the
existence of all (or the majority) of SMBHs, there are implied
constraints on the nature of dark matter. In particular, for
self-annihilating WIMPs, these particles must be captured at sufficent
rates into Pop III.1 protostars to allow protostellar growth to
$\sim10^5\:M_\odot$.

\section*{Acknowledgments}
We thank the following collaborators, especially in the SMBH and HUDF
groups, for their help in developing and testing the Pop III.1 seeding
model: G. Abramo; N. Banik; E. Blackman; K. Chanchaiworawit; O. Cray;
R. Ellis; F. Fontanot; K. Fridell; I. Georgiev; Y. Harikane; M.
Hayes; B. Keller; M. la Torre; N. Laporte; C. McKee; A. Nataranjan;
R. O'Reilly; B. O'Shea; A. Saxena; K. Tanaka; K. Topalakis; A.
Young. We also thank the following colleagues for helpful discussions:
B. Draine; X. Fan; K. Freese; C. Hill; C. Ilie; A. Lidz; G. Meynet;
D. Spergel; N. Yoshida. JCT acknowledges support from ERC Advanced
Grant 788829 (MSTAR) and the CCA Sabbatical Visiting Researcher
program.



\newpage

\begin{thebibliography}{00}


\bibitem{Aalbers24} Aalbers, J. et al. 2024, Physical Review D, Vol. 109, Issue 9, article id.092003

\bibitem{Abdurashidova22} Abdurashidova, Z. et al. 2022, ApJ, 925, 221

\bibitem{Abel02} Abel T., Bryan G. L., \& Norman M. L., 2002, Science, 295, 93

\bibitem{Ade19} Ade, P. et al. 2019, JCAP, Issue 02, article id. 056

\bibitem{Agazie23a} Agazie, G. et al. 2023, ApJ Lett., 951, L8

\bibitem{Agazie23b} Agazie, G. et al. 2023, ApJ Lett., 952, L37

\bibitem{Allys23} Allys, E. et al. 2023, PTEP, Vol. 2023, Issue 4, id.042F01

\bibitem{Amaro-Seoane23} Amaro-Seoane, P., et al. 2023, LRR, 26, 2

\bibitem{Baldassare15} Baldassare V. F., Reines A. E., Gallo E. \& Greene J. E. 2015, ApJ, 809, L14

\bibitem{Banik19} Banik N., Tan J. C., Monaco P., 2019, MNRAS, 483, 3592

\bibitem{Baumgardt17} Baumgardt, H. 2017, MNRAS, 464, 2174

\bibitem{Baumgarte99} Baumgarte T. W. \& Shapiro S. L. 1999, ApJ, 526, 941

\bibitem{Birchall20} Birchall, K. L., Watson, M. G., \& Aird, J. 2020, MNRAS, 492, 2268

\bibitem{Bogdan24} Bogd\'an, A., Goulding, A., Natarajan, P. et al. 2024, Nature Astronomy, Vol. 8, 126

\bibitem{Bouwens10} Bouwens, R. J., Illingworth, G. D., Oesch, P. A. et al. 2010, ApJL, 709, L133

\bibitem{Bromm02} Bromm V., Coppi P. S. \& Larson R. B., 2002, ApJ, 564, 23

\bibitem{Bromm03} Bromm V. \& Loeb A., 2003, ApJ, 596, 34

\bibitem{Buchert93} Buchert T. \& Ehlers J. 1993, MNRAS, 264, 375

\bibitem{Cammelli24} Cammelli, V., Monaco, P., Tan, J. C. et al. 2024, MNRAS, in press (arXiv:2407.09949) [CMT24]

\bibitem{Carr20} Carr, B. \& Kuhnel, F. 2020, Ann. Rev. Nucl. Part. Sci., 70, 355

\bibitem{Catelan95} Catelan P. 1995, MNRAS, 276, 115

\bibitem{Chandrasekhar64} Chandrasekhar S. 1964, ApJ, 140, 417

\bibitem{Chon16} Chon S., Hirano S., Hosokawa T. \& Yoshida N., 2016, ApJ, 832, 134

\bibitem{Ellis13} Ellis, R. S., McLure, R. J., Dunlop, J. S., et al. 2013, ApJL, 763, L7

\bibitem{Fontanot20} Fontanot, F., De Lucia, G., Hirschmann, M. et al. 2020, MNRAS, 496, 3943

\bibitem{Freese10} Freese K., Ilie C., Spolyar D., Valluri M., \& Bodenheimer P. 2010, ApJ, 716, 1397

\bibitem{Freitag06} Freitag, M., G\"urkan, M. A. \& Rasio, F. A. 2006, MNRAS, 368, 141

\bibitem{Gardiner24} Gardiner, E. C., Kelley, L. Z., Lemke, A.-M. \& Mitridate, A. 2024, ApJ, 965, 164

\bibitem{Greene20} Greene, J. E., Strader, J., Ho, L. S. 2020, ARA\&A, 58, 257

\bibitem{Greif06} Greif T. H. \& Bromm V., 2006 MNRAS, 373, 128

\bibitem{Gurkan04} G\"urkan, M. A., Freitag, M., \& Rasio, F. A. 2004, ApJ, 604, 632

\bibitem{Haberle24} H\"aberle, M., Neumayer, N., Seth, A. et al. 2024, Nature, Vol. 631, Issue 8020, 285

\bibitem{Haehnelt93} Haehnelt M. G. \& Rees M. J. 1993, MNRAS, 263, 168

\bibitem{Harikane23} Harikane, Y., Zhang, Y., Nakajima, K. et al. 2023, ApJ, 959, 39

\bibitem{Hayes24} Hayes, M., Tan, J. C., Ellis, R. et al. 2024, ApJL, 971, L16

\bibitem{Heger02} Heger, A. \& Woosley, S. E. 2002, ApJ, 567, 532

\bibitem{Heinrich17} Heinrich, C. H., Miranda, V. \& Hu, W. 2017, PRD, Vol. 95, Issue 2, id.023513

\bibitem{Hennawi06} Hennawi, J. F., Strauss, M. A., Oguri, M. et al. 2006, AJ, 131, 1

\bibitem{Hickox06} Hickox, R. C., Jones, C., Forman, W. R. et al. 2009, ApJ, 696, 891

\bibitem{Hirano14} Hirano, S., Hosokawa, T., Yoshida, N. et al. 2014, ApJ, 781, 60

\bibitem{Hosokawa11} Hosokawa T., Omukai K., Yoshida N. \& Yorke H. W., 2011, Science, 334, 1250

\bibitem{Ilie23} Ilie, C., Paulin, J., \& Freese, K. 2023, PNAS, Vol. 120, Issue 30, article id.e2305762120

\bibitem{Johnson06} Johnson, J. L. \& Bromm, V. 2006, MNRAS, 366, 247

\bibitem{Jungman96} Jungman, G., Kamionkowski, M. \& Griest, K. 1996, Phys. Rep., 267, 195

\bibitem{Kormendy13} Kormendy J. \& Ho L. C. 2013, ARA\&A, 51, 511

\bibitem{Kocevski23} Kocevski D. D., et al. 2023, ApJ, 954, L4

\bibitem{Latif16} Latif M. A. \& Schleicher D. R. G., 2016, A\&A, 585, 151

\bibitem{Latif22} Latif, M. A., Whalen, D. et al. 2022, Nature, Vol. 607, 7917, 48

\bibitem{Li24} Li J., Silverman, J. D., Shen, Y. et al. 2024, ApJ, sub. (arXiv:2403.00074)

\bibitem{Liepold24} Liepold, E. R. \& Ma, C. 2024, ApJ Lett., 971, L29

\bibitem{Lyu22} Lyu, J., Alberts, S., Rieke, G. H. \& Rujopakarn, W. 2022, ApJ, 941, 191

\bibitem{Maiolino24a} Mailolino, R., Scholtz, J., Witstok, J. et al. 2024, Nature, Vol. 627, 8002, 59

\bibitem{Maiolino24b} Maiolino, R., Scholtz, J., Curtis-Lake, E. et al. 2024, A\&A, 691, 145

\bibitem{Matsuoka24} Matsuoka, Y., Izumi, Takuma, T., Onoue, M. et al. 2024, ApJ, 965, L4

\bibitem{McKee08} McKee C. F. \& Tan J. C., 2008, ApJ, 681, 771

\bibitem{Millea18} Millea, M. \& Bouchet, F. 2018, A\&A, 617, A96

\bibitem{Miller15} Miller B. P., Gallo E., Greene J. E. et al. 2015, ApJ, 799, 98

\bibitem{Monaco13} Monaco P., Sefusatti E., Borgani S. et al. 2013, MNRAS, 433, 2389

\bibitem{Monaco02} Monaco P., Theuns T. \& Taffoni G. 2002, MNRAS, 331, 587

\bibitem{Moutarde91} Moutarde F., Alimi J.-M., Bouchet F. R., Pellat R., Ramani A., 1991, ApJ, 382, 377

\bibitem{Mummery24} Mummery, A. \& van Velzen, S. 2024, MNRAS, sub. (arXiv:2410.17087)

\bibitem{Mutlu-Pakdil16} Mutlu-Pakdil, B., Seigar, M. S. \& Davis, B. L. 2016, ApJ, 830, 117

\bibitem{Nakajima23} Nakajima, K., Ouchi, M., Isobe, Y. et al. 2023, ApJS, 269, 33

\bibitem{Natarajan09} Natarajan A., Tan J. C. \& O’Shea B. W. 2009, ApJ, 692, 574

\bibitem{Nguyen19} Nguyen D. D., et al., 2019, ApJ, 872, 104

\bibitem{Pacucci23} Pacucci, F., Nguyen, B., Carniani, S., Maiolino, R., Fan X. 2023, ApJ, 957, L3

\bibitem{Paxton13} Paxton, B., Cantiello, M., Arras, P. et al. 2013, ApJS, 208, 4

\bibitem{Rees78} Rees M. J., 1978, The Observatory, 98, 210

\bibitem{Reines15} Reines A. E. \& Volonteri M. 2015, ApJ, 813, 82

\bibitem{Reines16} Reines A. E. \& Comastri A., 2016, PASA, 33, 54

\bibitem{Rindler-Daller15} Rindler-Daller T., Montgomery M. H., Freese K. et al. 2015, ApJ, 799, 210

\bibitem{Rosdahl13} Rosdahl, J., Blaizot, J., Aubert, D. et al. 2013, MNRAS, 436, 2188

\bibitem{Schaeffer23} Schaeffer, T., Giri, S. K., Schneider, A. 2023, MNRAS, 526, 2942

\bibitem{Schleicher22} Schleicher, D. R. G., Reinoso, B., Latif, M. et al. 2022, MNRAS, 512, 6192

\bibitem{Shankar20} Shankar, F. et al. 2020, Nature Astronomy, 4, 282

\bibitem{Singh24} Singh, J. 2024, Ph.D. thesis, University of Trieste

\bibitem{Singh23} Singh, J., Monaco, P., Tan, J. C. 2023, MNRAS, 525, 969 [SMT23]

\bibitem{Smith12} Smith R. J., Iocco F., Glover S. et al. 2012, ApJ, 761, 154

\bibitem{Spolyar08} Spolyar D., Freese K. \& Gondolo P. 2008, Phys. Rev. Lett., 100, 051101

\bibitem{Stacy14} Stacy A., Pawlik A. H., Bromm V. \& Loeb A., 2014, MNRAS, 441, 822

\bibitem{Susa14} Susa H., Hasegawa K. \& Tominaga N., 2014, ApJ, 792, 32

\bibitem{Tan14} Tan, J. C., Beltran, M. T., Caselli, P. et al. 2014, Protostars \& Planets VI, p149

\bibitem{Tan04b} Tan J. C. \& Blackman E. G., 2004, ApJ, 603, 401

\bibitem{Tan04a} Tan J. C. \& McKee C. F., 2004, ApJ, 603, 383

\bibitem{Tan10} Tan, J. C., Smith, B. D., \& O'Shea, B. W. 2010, AIP Conf. Proc., Vol. 1294, 34

\bibitem{Tanaka17} Tanaka, K. E. I., Tan, J. C. \& Zhang, Y. 2017, ApJ, 835, 32

\bibitem{Ubler23} \"Ubler H., Maiolino, R., Curtis-Lake, E. et al., 2023, A\&A, 677, A145

\bibitem{Vika09} Vika M., Driver S. P., Graham A. W., Liske J., 2009, MNRAS, 400, 1451

\bibitem{Vogelsberger14} Vogelsberger M., et al. 2014, Nature, 509, 177

\bibitem{Volonteri21} Volonteri, M., Habouzit, M. \& Colpi, M. 2021, Nature Reviews Physics, 3, 732

\bibitem{Wang21} Wang, F., Jinyi, Y., Fan, X. et al. 2021, ApJ Letters, 907, L1

\bibitem{Weinberger17} Weinberger, R., Springel, V., Hernquist, L. et al. 2017, MNRAS, 465, 3291

\bibitem{Wise19} Wise, J. H., Regan, J. A., O’Shea, B. W. et al. 2019, Nature, 566, 7742

\bibitem{Wu22} Wu, Q. \& Shen, Y. 2022, ApJS, 263, 42

\bibitem{Yang21} Yang, J., Wang, F., Fan, X. et al. 2021, ApJ, 923, 262

\bibitem{Yue21} Yue, M., Fan, X., Yang, J. \& Wang, F. 2021, ApJ Lett., 921, L27

\bibitem{Yue23} Yue, M., Fan, X., Yang, J. \& Wang, F. 2023, AJ, 165, 191 

\bibitem{Zeltyn24} Zeltyn, G., Trakhtenbrot, B., Eracleous, M. et al. 2024, ApJ, 966, 85




\end{thebibliography}
\end{document}